# Feeling Anxious? Perceiving Anxiety in Tweets using Machine Learning

Dritjon Gruda[1], Souleiman Hasan[1]

[1] National University of Ireland Maynooth, School of Business, Maynooth, Ireland

**Abstract**

This study provides a predictive measurement tool to examine perceived anxiety from a longitudinal perspective, using a non-intrusive machine learning approach to scale human rating of anxiety in microblogs. Results suggest that our chosen machine learning approach depicts perceived user state-anxiety fluctuations over time, as well as mean trait anxiety. We further find a reverse relationship between perceived anxiety and outcomes such as social engagement and popularity. Implications on the individual, organizational, and societal levels are discussed.

# Introduction

Many studies in various disciplines have made use of individuals' digital footprints, interpreting uploaded information and making predictions about future behaviour based on behavioural residue (e.g. De Choudhury, Counts, & Horvitz, 2013; De Choudhury, Counts, Horvitz, & Hoff, 2014; Eichstaedt et al., 2015; Matz & Netzer, 2017; Settanni & Marengo, 2015). Behavioural residue is a trail of actions or deliberately placed markers which form identity claims (Gosling, Ko, Mannarelli, & Morris, 2002, p. 381). Several studies have used behavioural residue in digital footprints to assess and derive clues with regard to personality predictions (e.g. Azucar, Marengo, & Settanni, 2018; Li, Li, Hao, Guan, & Zhu, 2014; Qiu, Lin, Ramsay, & Yang, 2012) and other psychological characteristics (Kosinski, Matz, Gosling, Popov, & Stillwell, 2015; Settanni, Azucar, & Marengo, 2018). In this paper, we suggest examining anxiety in the naturally occurring setting (Qiu et al., 2012) of microblogs, using a machine learning approach.

Anxiety is a particularly interesting phenomenon to study, since anxiety occurs in both state and trait form. State-anxiety (Spielberger, Gorsuch, Lushene, Vagg, & Jacobs, 1983) is defined as "a temporary state influenced by the current situation where the respondent notes how he/she feels right now at this moment", whereas trait anxiety is defined as "a general propensity to be anxious where the respondent notes how he/she feels generally". Furthermore, a closer examination of the State-Trait Anxiety Inventory (STAI) reveals that state and trait anxiety differ mostly in terms of frequency of occurrence. Hence, while state anxiety measures how participants feel right now, trait anxiety captures the frequency of feeling anxious in general (i.e., not at all, frequently, a lot, etc.). For example, individuals are more likely to feel more anxious before certain events, e.g. an important presentation at work or an annual performance review meeting (i.e. state anxiety). However, in order to provide a full picture and reduce the likelihood of misrecognising state anxiety as trait anxiety (more prevalent in cross-sectional and self-report data), a longitudinal analysis is much more adequate.

To overcome this obstacle from a methodological point of view, we combine human zero-acquaintances ratings with a machine learning approach. Computational power allows us to evaluate a large amount of text input, and provides us with an unprecedented opportunity to track users' anxiety levels over time, a large limitation in previous studies (e.g., Qiu et al., 2012). Our machine learning approach paired with data crawling techniques allows us to collect a large enough dataset (Kosinski, Stillwell, & Graepel, 2013; Wald, Khoshgoftaar, Napolitano, & Sumner, 2012), which we also overcomes small sample size biases and self-selection sampling in previous studies (Settanni & Marengo, 2015). Proper training and analysis could reduce the effect of the machine learning error through representative and multi-rater labelling, founded modelling to reduce overfitting (e.g. through cross-validation), along with aggregate analysis to offset the random machine prediction error.

In order to test our presented approach and its implications regarding future outcomes, we relate the degree of perceived user anxiety, both current (i.e. state anxiety) and frequency of expression (i.e. trait anxiety) to the degree of popularity and social engagement of Tweeters, i.e. the number of followers and number of users one follows, respectively. Since these two outcome variables are measured only at the point of data collection and anxiety as detected in tweets is measured over time, we can make predictions about future popularity and social engagement using anxiety as a predictor. Related research in the information systems literature (Quercia, Kosinski, Stillwell, & Crowcroft, 2011) found that predictive statements can be made regarding traits based on just a handful of criteria. These include publicly available factors, namely the amount of following, followers and listed counts on Twitter. For example, popular users (those with a large group of followers) and listeners (those who follow many others) score low on neuroticism. Since individuals who score high on neuroticism tend to withdraw to themselves, especially in stressful situations (Lee-Baggley, Preece, & DeLongis, 2005), these findings also translate into real life, where neuroticism is a useful predictor of the number of friends in life (Swickert, Rosentreter, Hittner, & Mushrush, 2002) as well as Facebook contacts (Golbeck, Robles, & Turner, 2011). Therefore, similarly to previous research, we use anxiety detected in tweets to make predictions about individuals' degree of popularity and social engagement in life. Although implied in previous research (Quercia et al., 2011), to the best of our knowledge, the presented study is the first to do so using very high volumes of data over time to derive useful implications based on the detection of anxiety in microblogs.

**Measuring Anxiety: Self-Report vs. Observer-Rated**

Anxiety can be measured in various ways, including self-report, physiological, and behavioural measures (Eysenck, 2000), yet, concordance between these measures remains low (Newton & Contrada, 1992). Most likely, this is due to cognitive biases associated with anxiety (Eysenck, 2000), in particular comparing self-report with observer-rated measures.

For example, since highly anxious individuals (i.e., high trait anxiety) tend to interpret perceived information in an exaggerated threatening fashion, these individuals usually report higher self-report anxiety than observer-rated physiological or behavioural measures would indicate. Those who try to repress their anxiety levels, i.e. repressors, aim to reduce the threat level of new information (Eysenck, 2000). In this case, self-report anxiety levels are more likely to be lower than observer-rated measures. Finally, individuals who score low on *trait* anxiety are not prone to the same cognitive biases. Hence, their self-report anxiety is comparable to observer-rated anxiety. Due to the existence of these above-mentioned cognitive biases, as well as the human tendency to generally answer self-report survey questions favourably (i.e. social desireability, Fisher, 1993), it would be much more effective to monitor and track observer-rated anxiety levels of individuals rather than self-report anxiety.

Nonetheless, observer-rated measurements are numerous, which hinders a direct comparison of results. For example, based on a sample of outpatients, Schat et al. (2017) found an overall strong and positive correlation between self-report and observer-rated measures of anxiety severity. However, in 12.6% of cases patients reported higher self-report anxiety than observer-rated anxiety (p. 9), while in 12.2% of cases higher observer-rated than self-report anxiety scores were reported. These results, however, likely provide a skewed picture of overall anxiety levels and likely are not representative of the general population. For example, patient questions about anxiety would include a measurement of phobias, reduced sleep or muscular tension (BAS; observer-rated), and faintness, chest pains, and trouble breathing (BSI; self-report). While such questions are important in assessing a patient's anxious tendencies and determine treatment options (Dorz, Borgherini, Conforti, Scarso, & Magni, 2004), the same factors could be confounding results in case of zero-acquaintance raters evaluating anxiety based on online data.

### Measuring Anxiety Online

In order to create a predictive model of anxiety in the general population, we base the measurement of perceived anxiety on zero-acquaintances, who rate perceived anxiety levels in others by rating micro-blogs. To do so we firstly need to address whether it is even possible to reliably do so. For example, De Choudhury et al. (2013) accurately measured users' depression levels (self-report/observer-rated scores of depression) by analysing user tweets. In addition, the respective authors could also make accurate predictions about the future onset of depression. Another study (Qiu et al., 2012) found that aggregated observer ratings of neuroticism (and agreeableness) in participants' tweets were significantly correlated to self-report personality ratings. This result is in line with previous studies (e.g. Holleran & Mehl, 2008).

Previous research (Youyou, Kosinski, & Stillwell, 2015) showed that even the mere examination of Facebook likes can serve to make accurate predictions with regard to users' personality, showed higher interrater agreement, and were better predictors of other life outcomes such as physical health. Indeed, computer predictions were almost as accurate as participants' spouses in the assessment of participants' personality. Hence, the greater the amount of behavioural residue, i.e. the more available data, the higher the accuracy of provided ratings (Settanni et al., 2018).

Likewise, the same principle applies to the number of raters per behavioural clue in linguistic form. Complete strangers (i.e., zero-acquaintance raters) can make accurate personality judgments after observing participants' behaviour (Funder, Kolar, & Blackman, 1995; Macrae & Quadflieg, 2010; Ready, Clark, Watson, & Westerhouse, 2000). What makes these zero-acquaintances ratings accurate is a connection between "the presence of deliberate or unintentional self-expressions and, importantly, communication of some form of information to others" (Tskhay & Rule, 2014). In addition, we do not ask our raters to rate tweets on anxiety specifically, but repurpose a well-established scale (STAI; Spielberger et al., 1983) to evaluate each tweet on anxiety. We do so instead of using linguistic software such as LIWC (Pennebaker, Francis, & Booth, 2001). LIWC, although highly beneficial if the goal is linguistic usage of words, overlooks more subtle clues such as higher-order semantic cues (Gill & Oberlander, 2002; Hirsh & Peterson, 2009). We have conducted an empirical comparison with LIWC as explained in detail in the methodology section (Step 6). The comparison shows that the machine learning approach outperforms LIWC in terms of regression to human rating. Nonetheless, in the absence of any human labelling that can form a basis for machine learning, unsupervised methods, which are based on dictionary word counts, such as LIWC, can form an alternative approach.

However, the number of raters is important. For example, Qiu et al. (2012) found that a single observer cannot make accurate predictions when it comes to determining participants' personality traits in micro-blogs (p. 713). Therefore, multiple raters are needed to achieve a respectable degree of judgment accuracy. Yet, observers likely rely on similar cues when providing their ratings, as long as inter-observer agreement is high (Graham & Gosling, 2012; Vazire & Mehl, 2008).

## Measuring Anxiety in Tweets

In this study, we focus on examining anxiety in micro-blogs, namely tweets. Twitter is the world's most popular microblogging platform, with over 328 million active users (Statista, 2017), with 100 million daily users (Aslam, 2017) who produce on average 6,000 tweets (140-character text messages) per second, or 500 million tweets per day. Twitter users also can subscribe to what other users post, known as followers. Microblogging is quick, short, and mostly captures what is going on at any particular moment (Oulasvirta, Lehtonen, Kurvinen, & Raento, 2010). Twitter users' news feed therefore provides a good fit for analyses in the social sciences, as tweets mostly capture users' thoughts, feelings, and conversations at a particular moment in time (Naaman, Boase, & Lai, 2010). One of the most important distinctions between Twitter and Facebook is the degree of anonymity. While Twitter allows users to create a pseudonymous profile to share opinions (Hughes, Rowe, Batey, & Lee, 2012; Kwak, Lee, Park, & Moon, 2010), Facebook requires its users to enter their real name and post actual user information, in order to find friends and interact with 'people you may know' (Huberman, Romero, & Wu, 2009). Tweets (for the most part) are public by default while on Facebook, the default setting of posts is set to private, as information is shared primarily with friends or familiar people in users' social circles. Another argument could be that Twitter and Facebook users also differ with regard to their personality. However, two recent meta-analyses found no difference based on social media platform type (public vs. private) with regard to the accuracy of personality prediction (Azucar et al., 2018; Settanni et al., 2018).

In our study, participants' behavioural residue is composed of tweet content and the way participants use words and phrases, rated by human raters and scaled by using a machine learning algorithm. We use Twitter because Twitter data is publicly available and non-intrusive. Doing so allows us to capture naturally occurring expressions of a wide and representative set of users (Qiu et al., 2012), which is in contrast to laboratory studies in which participants would not necessarily express themselves naturally (e.g. Kanagawa, Cross, & Markus, 2001). In addition, microblog data include a high degree of social interaction with others. Users oftentimes engage in debates with others on social media platform, and share like-minded posts with others. All these aspects contribute to the accumulation of linguistic and behavioural cues and the overall expression of oneself, in a way that was until recently very difficult to analyse (Golbeck et al., 2011; Yee, Ducheneaut, Nelson, & Likarish, 2011).

## Methodology

Our methodology is based on content analysis. Hence, for the analysis to take place anxiety scores are needed per tweet. Manual scoring is not feasible if the content is to span across many tweets (i.e. millions), and over large time spans (i.e. years). For this purpose, we propose a methodology that is hybrid between human and machine rating, where human scaling for a dataset of 600 tweets is used to train the machine how to score tweets. Machine learning is then used in a subsequent stage to score millions of tweets resulting in a comprehensive data analysis. In Figure 1 we provide an overview and illustrate our research methodology.

-------------------------------------
Insert Figure 1 about here
-------------------------------------

The three phases of the methodology include:

A. Labelled data preparation: a set of 600 tweets (US tweeters) are manually rated by 604 participants (US participants only) on anxiety (Marteau & Bekker, 1992), each human rater rating 5 tweets on average, which resulted in 3,020 observations.

B. Machine learning: features are extracted from tweets' textual content to learn how human raters rated, resulting in learned machine models.

C. Anxiety scoring and data analysis: a final dataset of about 3,33 million public tweets by 1,418 users, over an average of 18-months timeline, is prepared and scored using machine learning. Data analysis takes place on this dataset as reported in the results section. The subset of 1,418 users has been randomly selected for practical reasons to handle the resulting volume of data, and is a large number sufficient for statistical analysis. The number of 3.33 million tweets is a result of retrieving what the Twitter API allows for those users. Since the Twitter API allows the retrieval of up to 3,200 tweets for any user, the time span may vary per user's tweeting activity. On average, the time span is around 18 months per user.

**Step 1: Search for Tweets**

The purpose of this step is to retrieve a set of tweets using the keywords "work" and "feeling" to form the basis of a human-labelled dataset. The Twitter API is called with the standard search API method and the mentioned keywords to retrieve tweets from the live stream. This was done over the period of two days in December 2017. Details and limits imposed by the Twitter API can be found in its online reference[1].

---
[1] https://developer.twitter.com/en/docs/tweets/search/api-reference/get-search-tweets

This step resulted in a dataset of 10,510 tweets, made by 10,386 users, each of which mentions the search terms. 60% of these tweets are retweets, and 9% are replies. Due to retweets and replies, textual content might repeat although tweet objects are different. In fact, around 46% of the tweets have unique textual content.

**Step 2: Manually Select a Subset of Tweets**

A subset of tweets is selected to reflect a balanced[2] normal distribution of anxiety so that subsequent machine learning avoids biases due to imbalanced data. 600 unique tweets have been selected by selecting non-retweets and non-replies, removing links, and removing characters that are not correctly appearing. The Twitter search API, originally used to retrieve the tweets, may not include the geographical location of each user. This is because the user might not provide this information. Thus, we have selected tweets with English content of users whose location or time zone are explicitly mentioned and conform to the USA to avoid cultural bias.

**Step 3: Rate Anxiety by Human Raters**

Perceived anxiety was measured with the short version of the traditional full scale State-Trait Anxiety Inventory (STAI; Spielberger et al., 1983). The final scale is composed of 6 items on a four-point scale. For example, a tweet with the content "At work feeling terrible 😣" is scored 3.4 out of 4 on average by six raters, representing high anxiety. In another instance, a tweet with the contnet "Was feeling myself at work today" is scored 1.5 on average by 5 raters, indicating low anxiety.

-------------------------------------
Insert Figure 2 about here
-------------------------------------

---

[2] Balance of the selected set has been confirmed before moving on to final labelling, by refining Steps 1 and 2. We also conducted two pre-tests of 50 tweets and 50 raters each, with each rater rating more than one tweet. This was done in order to assure interrater agreemets and distribution of anxiety.

Due to the longitudinal nature of our study design, this scale is sufficient even though it only asks participants to evaluate how the person tweeting (i.e. the tweeter) feels in one tweet at a time, as trait anxiety can be measured at data analysis by consider multiple tweets over time. After recoding reversed items, the total state-anxiety score is derived by summing all items ($\alpha = .90$). Interrater agreement was calculated based on Intra-Class Correlations (ICC; Koch 1982) using STATA 15. Using the full labelled dataset of 600 tweets, we derived an acceptable ICC (= .69, SE=.015; (95% CI = [.66, .72]).

Figure 2 illustrates a dot plot of individual human rating per tweet for a random set of tweets from the 600 dataset. It shows a tendency for the dots to cluster for each tweet, which reflects a high correlation between raters on how they judge a particular tweet. The interclass correlation coefficient (ICC) of .69 summarizes this pattern for the entire dataset. We would argue that from a methodological perspective, it is important that the labelling process accounts for multiple labels by different raters, and that raters' ratings correlate with each other. This increases the trust in the labelled data, which forms the basis of machine learning to try to mimic human ratings. If human raters do not agree on their rating, one would not expect the machine to be able to accurately learn how to score the data.

**Step 4: Calculate Mean Anxiety Scores**

Multiple scores of multiple raters of the same tweet are averaged into one score per tweet, which is then used for machine learning. Scores were found to be balanced in the range of 1-4, which reduces bias in machine learning in later stages.

**Step 5: Feature Extraction**

Next we need to ensure that machine scores mimic human scores, i.e. to allow the machine to replicate the human rater role, and then to rate many more tweets automatically. Features are extracted from the text of each tweet of the 600 labelled tweets. This is done since machine learning algorithms are generally not suited to work directly on words and sentences, but rather on numerical vectors, or features, of text.

The first type of feature is the *semantic embedding vector*, which is the mean of multiple vectors that map each word in the tweet to a distributional vector of 300 dimensions. This accounts for similarity between words such as synonyms, which typically have vectors closer than those of unrelated words. Here, every tweet becomes a vector of 300 dimensions with each component having continuous negative or positive values. Such a vector summarizes the meaning of the whole tweet. Two similar tweets in meaning would have close vectors in the mathematical sense (e.g. Cosine Similarity) rather than semantically different tweets. Words-to-vectors mapping is based on the deep neural network learning GloVe (Pennington, Socher, & Manning, 2014) embedding space built from the Common Crawl Web Data (42 Billion tokens, 1.9M vocab).

The second type of feature are *words and emojis occurrence vectors*. For example, a tweet that has the content "no better feeling 😏 😏 😏" is converted into a vector of the form [no: 1, better: 1, feeling : 1, 😏 : 3 ] to reflect word and emojis occurrence. Such a vector will have entries with zeros for absent words. It will also have entries for bigrams such as no-better, better-feeling, 😏😏, etc. Tweets with the same words or emoji occurrences will be considered close when the machine learning tries to fit a model for the tweets and their scores.

**Step 6: Regression Models Machine Learning**

The training data, consisting of the 600 tweets along with their mean anxiety scores assigned by human raters, have been used to train two models. Each model works on one type of features (i.e. semantic embedding vectors, and words and emoji occurrence vectors). This stage is experimental in a data science pipeline, in the sense that the models which perform best when trained on a part of the labelled data, and tested on the rest, should be adopted.

Among the various models we have experimented with, the Bayesian Ridge Regression (MacKay, 1992) performed best. Hence, this is a supervised method, but one can also perform an unsupervised method, using the dictionary-based LIWC for instance, in the absence of any training data. We draw such a comparison at the end of this Step 6.

The problem is a regression problem since the predicted value, i.e. the anxiety score, is continuous. Bayesian Ridge Regression is based on our implementation of linear Ridge regression and follows the Python Scikit Learn library. If the features are X, and the anxiety scores are Y, Ridge Linear regression looks for W, which minimizes the loss function:

$$\|Y - XW\|_2^2 + \alpha \|W\|_2^2$$

Where $\| \ \|_2^2$ denotes the L2 norm and $\alpha$ is the weight of the regularization term. The Ridge regression does not fit a linear model that has least squares only (i.e. $\|Y - XW\|_2^2$) but regularizes based on the norm of W (i.e. $\|W\|_2^2$). This makes it less likely to over-fit, and thus generalizes better on unseen tweets. The Bayesian Ridge Regression assumes prior distributions of the hyper parameters, which are jointly estimated during the training process. Thus, given the theoretical characteristics of the learning model, as well as its empirical validation discussed below, one can reasonably trust the model's scoring on unseen tweets.

We fit two regression models to the two sets of features and the corresponding human-given anxiety scores. The two models are learned based on the two types of features. At prediction time of anxiety of non-labelled tweets, the mean of the two predicted scores of the two models is taken.

The learning and testing process has been performed through a six-fold cross-validation, in which the 600 tweets are split into 6 subsets of 100 tweets each, and every time one of the subsets is used for testing while the rest is used for the training. The performance is the average performance of the six folds. A good learned model should predict scores for the tweets that would correlate with the scores of the human raters.

------------------------------------
Insert Figure 3 about here
------------------------------------

Performance measures (i.e., goodness of fit $R^2$ = .49 and Spearman Rank Correlation = .69) suggest that if a human would rank tweet A more anxious than tweet B, then the machine would likely do the same with a spearman correlation of .69, which is also illustrated in Figure 3. Performance is also summarized in the Root-Mean-Square Error (RMSE) = .52 which represents the root of the mean of squared differences between the machine-predicted scores and the human-assigned scores. The value .52 is small with respect to the range [1-4] of the scale.

------------------------------------
Insert Figure 4 about here
------------------------------------

***Error is smaller when considering multiple tweets***. It is noteworthy that the performance and error discussion above is applicable when considering individual tweets only. However, if the downstream data analysis aggregates scores of multiple tweets, the error of the machine learning starts to cancel out as some of the error would be negative and some would be positive. This makes data analysis with aggregate (e.g. mean) scores more reliable and prone to smaller error, since the machine learning error is almost normally distributed with a mean of 0 as shown in Figure 4.

------------------------------------

Insert Figure 5 about here

-------------------------------------

***To compare the performance with the LIWC software***, we have run LIWC on the same tweets that the machine learning was performed on, and of which we know the human scoring. LIWC as a software is based on a dictionary and relies on counting words, in an unsupervised manner. Words can be categorized under categories such as anxiety, positive emotions or negative emotions. In this study, we used the most recent LIWC2015 dictionary. This dictionary uses 116 words or word roots to account for anxiety (ANX category), if such words appear in a text or tweet.

Like other dictionary-based methods, this has the advantage of not relying on training and being an unsupervised method. On the other hand, the ability of LIWC is limited by the words defined in its dictionary for specific categories, such as anxiety. For example, a tweet with the content "*At work feeling terrible* 😔" has been labelled by human raters and machine learning to contain significant anxiety of around 3 out of 4 on the used scale. Nonetheless, LIWC found 0 words of its anxiety category in this tweet, and thus labelled it as zero-anxiety tweet.

The problem stems from three main reasons: (1) the dictionary may not cover all possible words that can describe a category like anxiety, (2) anxiety may be conveyed through a complex sentence that relies on an interaction between several words that together, and no single word alone, can describe the feeling, and (3) tweets are generally short text and may contain misspelled words or non-standard language which further challenges dictionary-based methods. These limitations could be remedied by machine learning from human-labelled data, given that such training data is available. More complex features than bags-of-words could help, by accounting for groups of words (n-grams), and the use of embedding semantic models, which can learn meanings such as relations between synonyms, using neural networks trained on large corpora of English-in-use, as the models used in this article.

Since the ANX (anxiety) category in LIWC yielded 0% in most tweets, and to devise a comparison with the machine-learning model, we turned to the emotion categories in LIWC since they cover more words and can yield a score for tweets. While these do not represent anxiety directly, we derive here a sentiment measure from emotion as a proxy for anxiety to enable the comparison. Previous studies have also shown negative/positive correlations between anxiety and positive/negative emotions respectively (Pekrun, Goetz, Frenzel, Barchfeld, & Perry, 2011; Watson, Clark, & Carey, 1988). We derived the sentiment score measure from the positive and negative emotion categories and normalized the range to [1-4], for comparison. The derived sentiment score is equal to (negative emotions *negemo* <u>minus</u> positive emotions *posemo*) normalized to [0-1], multiplied by 3, and finally added to 1:

$$SENTIMENT\ SCORE_{LIWC} = negemo - posemo \in [-1, +1]$$

$Normalized\ SENTIMENT\ SCORE_{LIWC}$
$$= 3 \times \frac{SENTIMENT\ SCORE_{LIWC} - Min(SENTIMENT\ SCORE_{LIWC})}{Max(SENTIMENT\ SCORE_{LIWC}) - Min(SENTIMENT\ SCORE_{LIWC})} + 1 \in [1,4]$$

Performance measures for LIWC, as shown in Figure 5, are lower than measures of the machine learning model: goodness of fit $R^2$ = .21 and Spearman Rank Correlation = .49, compared to $R^2$ = .49 and Spearman Rank Correlation = .69 in our machine learning model, respectively. The Root-Mean-Square Error (RMSE) equals .63 and is greater than the score of .52 in the machine learning model. Since LIWC counts words in text may sometimes find zero words from its dictionary to categorize under positive or negative emotions. This shortcoming results in a sentiment score of 2.5 (neither low nor high), which is seen as a horizontal line pattern in Figure 5. Overall, the machine learning model outperforms the LIWC approach, as its features are derived from more complex semantic models learned through neural network embeddings. The machine learning model also accounts for multiple word frequencies (bi-grams) and a larger set of emojis. However, dictionary and lexicon-based approaches such as LIWC, remain useful in applications with an absence of any training data as they run in an unsupervised manner.

**Step 7: Select Users Subset**

From the 10,386 users found in Step 1, we select a subset of users which we consider for preparing the final large dataset of tweets. We randomly selected 1,418 users for practical considerations of potential data volume and limits of the Twitter API.

**Step 8: Retrieve Users Timelines**

For each user selected in Step 7, we use the Twitter user timelines API [3] to retrieve each user's tweets from the past up to the day of collection. The API allows a maximum of 3,200 tweets per user to be retrieved, which in many cases spans back over several months or even years of user's activity. The result of this step is a set of 3,330, 387 public tweets, belonging to 1,418 users. The average number of tweets per user is 2,348, with 52% of tweets being retweets and 17% being replies. The dataset forms a non-invasive longitudinal study with an average time span of 18 months-worth of tweets per user.

**Step 9: Machine Learning Scoring**

The machine learning models resulting from Step 6 are now used to score the 3.33 million tweets resulting from Step 8. This results in a large dataset of scored tweets on the anxiety scale (M = 2.34, SD = .36). A very small amount of these (around .23%) are scored by the machine outside of the [1-4] range due to extrapolation. Due to the small percentage of those scores outside the range, we decided to ignore this data in our subsequent data analysis.

---

[3] https://developer.twitter.com/en/docs/tweets/timelines/api-reference/get-statuses-user_timeline.html

**Step 10: Data Analysis**

The final dataset is subjected to data analysis in order to draw conclusions including, for instance, the filtering of retweets and replies if needed, visualizing temporal user anxiety, the aggregation and mean scores, and conducting prediction models regarding users popularity and social engagement with anxiety scores. Popularity and social engagement are static variables in our dataset, i.e. they both were measured only at the point of data collection. That is due to a limitation in the data provided by Twitter. Although up to 3,200 tweets can be collected by user, which can reach back in time several months or even years, users' number of followers and number of users' followed are determined at the point of data collection and are not displayed in a dynamic manner. Hence, since these two outcome variables are static and anxiety as detected in tweets is measured over time, we can make predictions about future popularity and social engagement using anxiety as a predictor. A reverse causal relationship, although possible, cannot be accounted for in this study.

**Measurements**

*Anxiety* was measured with the short version of the traditional full scale (State-Trait Anxiety Inventory (STAI; Spielberger et al., 1983). Since the design of our study includes multiple ratings of several tweets per participant, we chose an abbreviated format of the STAI (Marteau & Bekker, 1992) composed of 6 items on a four-point scale, with 1 = "Not at all" and 4 = "Very much". For example, a tweet with the content "At work feeling terrible 😣" is scored 3.4 out of 4 on average by 6 raters, representing high state anxiety. In another instance, the tweet with the content "Was feeling myself at work today" is scored 1.5 on average by 5 raters, indicating low anxiety.

***Social engagement*** was measured by the number of people a tweeter follows, i.e. following count, whereas ***popularity*** was measured by considering the number of followers a Twitteruser has, i.e. followers count. Results are detailed in the results section.

**Results**

***Correlations*** among variables are listed in Table 1:

-------------------------------------
Insert Table 1 about here
-------------------------------------

***Data Visualization.*** In order to provide data visualization we randomly looked at anxiety scores of four random users, over time. In the figure below we highlight one example, a given time period of one week just before Christmas (from 15$^{th}$ – 23$^{rd}$ December). An example overview of four random tweeters in our database is shown in Figure 6. We observe for example that tweeter #1 and tweeter #3 both have fluctuations of state anxiety scores as expressed in their tweets; however, tweeter #1 has a higher mean (i.e. state) anxiety as opposed to tweeter #3.

-------------------------------------
Insert Figure 6 about here
-------------------------------------

***Regressions.*** In our model, we examined whether anxiety scores can predict both follower count (number of followers) as well as following count (number of people a tweeter follows). An initial examination of collected data points showed that we would capture approximately 82% of all our data if we were to consider time periods of 12 months (i.e. 2017), or 69% considering a period of 6 months (i.e. second half of 2017). Given our complete dataset of over three million observations, with observations spanning from 2 – 3,204 tweets per user, going back further than one year would yield a somewhat unbalanced dataset. Therefore, in Table 2 we report results based on the time periods 6 months and 12 months.

As explained above, each tweet was evaluated individually on the expression of state anxiety, combined to form an overall trait anxiety score per user. Since some tweeters tweet more often than others at any given time, total tweet count per user needs to be considered in the analysis and visualization of our data.

-------------------------------------
Insert Table 2 about here
-------------------------------------

The aggregated data by user over time (n = 1322) show that anxiety scores significantly predict both social engagement (i.e. following count) as well as popularity (follower count). The effects are quite strong, ranging from -.39 to -.77. Looking at the actual model fit, it is evident that our measure of perceived anxiety is a better predictor for social engagement over time (i.e. the number of other users a tweeter follows).

**Discussion**

We presented a tool that can detect the onset of significant changes in individuals' expressed anxiety over time. Since anxiety is not always evident, especially if anxiety levels remain high for a longer period (induced by a non-specific event, i.e. trait anxiety) it would become difficult for individuals to recognize the onset of anxiety, or the association between high anxiety levels when encountering a specific person or situation. However, due to the behavioural residue that is left behind in linguistic form, we can draw conclusions regarding how individuals feel at a given time. Some statements are clearer than others of course; for example, "I feel so anxious at work" much more reflects high anxiety levels than "Just not feeling work today". However, we are confident that these statements have been evaluated reliably, as interrater agreement was quite high (ICC = .69).

It is important to distinguish that in this study we do not measure actual experienced user anxiety, or short, anxiety. Instead, in this study we account for observer-rated perceived user anxiety, based on ratings of perceived anxiety provided by zero-acquaintances. We do not match perceptions of anxiety with actual user anxiety.

However, the presented approach and manner of data collection does allow us to measure perceived user anxiety in a non-invasive, unobtrusive and naturally occurring manner. In addition, our approach allows us to predict other outcomes as well. Based on previous work by Quercia et al. (2011), we analysed our data with the goal of predicting future outcomes, such as popularity and social engagement. Quercia et al. (2011) found that based on just a few Twitter profile characteristics, namely the following and followers count (and list count), one can predict tweeters' personality traits "with a root-mean-squared error below .88" (p. 183). Although quite impressive, their research design was only cross-sectional in nature, and does not account for changes over time. One could argue, since the Big-5 personality traits are quite stable over time (Goldberg et al., 2006) a longitudinal design might not be necessary. However, anxiety has both state and trait properties, which requires observation over time. We aimed to predict the same outcomes, using a time-sensitive, longitudinal measure of anxiety, accounting for anxiety intensity (state) and frequency (trait).

Interestingly, we find that perceived anxiety significantly and negatively related to both social engagement (following count) as well as popularity (follower count). These findings are in line with previous research (Quercia et al., 2011). The effect sizes are quite large, which provides confidence that a strong relationship between anxiety and online perception exists. It seems individuals who express a high degree of trait anxiety are less likely to follow others (social engagement), and are less often followed by others (popularity). Put differently the more popular a person is (on Twitter), the less anxiously rated statements they express or tweet, over time. Anxious individuals, in particular those who score high on trait anxiety, would be more inclined not to follow others, or accumulate a large follower community. In terms of practical implications of these findings, it is likely that highly anxious individuals potentially limit their social engagement with others. They are likely to do so as they would worry about being perceived negatively by others. Not wanting to show that one is anxious would lead to less proactive social engagement with others, and subsequently lower popularity. This negative view of self could even lead to increased experienced anxiety over time. Such feelings are likely to be embedded in the behavioural residue such as individuals' writing, expressions and written social media interactions. While elevated state anxiety is unlikely to cause such change of perceptions, due to the event-specific short duration, sustained elevated state anxiety at the workplace could potentially lead to social exclusion and even hinder career and management opportunities (Park, Woo, Park, Kyea, & Yang, 2017).

The interplay between perceived state and trait anxiety we found is vital and relates to two aspects in this study. Firstly, the method is intrinsically fine-grained, which allows the investigation of individual users' expressed anxiety level. In fact, aggregate concepts such as trait anxiety becomes a downstream data analysis step, as concepts of longitudinal study and frequencies are mathematical properties of the state anxiety time-series. In this study, we considered the mean state anxiety score as trait anxiety and regressed this factor onto social engagement and popularity at the aggregated level. Secondly, there is also a methodological reason behind this step, which lies in the small error embedded in the methodology, due to the machine learning limitation in exactly mimicking human raters. Such error, while small, can be unsatisfying when looking at an individual tweet. Nonetheless, as discussed in the methodology section, taking aggregates such as the mean, and hence trait anxiety, allows the error to cancel out due to its distribution. Thus, the trait anxiety score is a rather reliable measure of a user from a methodological perspective.

From a methodological perspective, the hybrid human-machine approach to anxiety labelling of tweets suggests that the sample of observations can be scaled vastly in multiple dimensions. For example, the temporal dimension of tweets conforming to a longitudinal extension, and the subjects' dimensions (i.e. tweeters) which, in principle, can be extended to a large proportion of the whole population on Twitter, spanning cultures, geographical zones, etc. Naturally, this would require additional computational power, and Big Data techniques such as parallel computing can come to the rescue. From a statistical perspective, testing hypotheses would lead to higher statistical power compared to a non-machine-based approach. However, we do not claim to match perceptions of anxiety with actual experienced anxiety, as we would need to ask all tweeters in our database to provide us with self-reports on their experienced anxiety levels during each tweet; a task virtually impossible to accomplish. This would also lead to the analysis of data in a non-natural environment, i.e. being taken out of the moment every time you tweet, so you can fill out an anxiety scale. Reflecting on current anxiety levels would likely also influence the tweet itself.

**Implications**

There are certain practical implications, which we can derive from our study. Our study looked at microblogs as the content to analyse expression of anxiety. Microblogs are pervasive in many social environments but also in professional contexts. For instance, team collaboration tools such as Slack[4] often include channels of messages to allow team members to communicate on certain project or topic. Automatic anxiety scoring of individuals and teams can help monitor the workers wellbeing and prevent situations of excessive anxiety that can lead to burnouts affecting the workers and projects alike. At the very least, using this tool employers would be less likely to foster an anxiety loaded work place.

---

[4] https://slack.com/

Of course, as is the case with any data collection tools (Azucar et al., 2018), this tool could also be exploited and misused. With recent data sharing incidents involving Facebook (Levin, 2017) and the possible dissemination of political propaganda to impressionable social media users (Cadwalladr, 2017; Confessore & Hakim, 2017), the possibilities of misconduct are numerous. Policy makers and the public should be made aware that such techniques are already being applied and might need to be restricted in usage, in order to protect individual privacy rights. On an organisational level, the monitoring of individual employees' emotional states, such as their state anxiety, could potentially also pose privacy violations, especially if companies exploit such knowledge at the expense of the employees.

Another possible implication of abusing the presented approach revolves around the stigma of anxiety. For example, singling out employees who have high trait anxiety, could lead to lower employee self-esteem, social rejection and reluctance to seek appropriate treatment (Barney, Griffiths, Jorm, & Christensen, 2006). In certain cases, employees might think that their elevated anxiety levels are only temporary, not severe enough to seek help or they misjudge the toll their elevated anxiety levels are taking on their health and well-being. Research on the issue of stigma of workplace health issues such as anxiety and depression has found that oftentimes the experienced stigma can be as negative as the mental health issue itself (Overton & Medina, 2008). In addition, should employees in question become known in the organization, it could lead to increased stereotyping and discrimination (Szeto & Dobson, 2010) with additional negative health consequences for the stigmatized employees as well as friends or close colleagues (Corrigan, 2004; Hinshaw, 2009; Thornicroft, Rose, Kassam, & Sartorius, 2007).

The findings of this study point towards the use of aggregate scores as they are more accurate than individual scores. That is due to the normal distribution of the error, which cancels out when many scores are aggregated. Thus, profiling based on anxiety scores may not be as accurate at the individual level, let alone the ethical issues such an approach implies. Using such a machine learning model to produce aggregate scores, is more accurate. Consequently, our approach could reflect the wellbeing of individuals or groups of individuals, which would allow the implementation of measures, such as organizational or governmental policies to be implemented to address and hopefully reduce increased anxiety. We encourage scholars to study the impact of the presented technology and its application in and out of the workplace in greater detail in future studies.

**Limitations**

Firstly, one of the limitations of this study is that the ratings provided by our zero-acquaintances, do not actually measure felt anxiety by the tweeter, but rather perceived anxiety. This has been raised as an in issue in previous work as well (Qiu et al., 2012). In order to make accurate predictions with regard to actual user anxiety further validation of our approach is needed. One potential validation would constitute combining the presented approach with other longitudinally assessed repeated self-report measures of anxiety in a group of participants. Participants' would provide daily self-report assessments on anxiety, while their social media communication could be examined using the presented approach. Doing so would allow researchers to compare the accuracy and validity of the presented method.

Secondly, an argument could be made that online behaviour greatly differs from offline behaviour, and hence deriving behavioural residue from online sources is not feasible in explaining relationships to psychological characteristics. However, the contrary has been found (Roberts, Kuncel, Shiner, Caspi, & Goldberg, 2007). Social media users seem to view their online profiles as an extension of the self, rather than a separate entity (Back et al., 2010; Seidman, 2013). Not only does online behaviour relate strongly to psychological characteristics, the accuracy to predict characteristics based on online data is becoming stronger over time, as data mining techniques become more sophisticated using machine learning (Azucar et al., 2018; Kosinski, Wang, Lakkaraju, & Leskovec, 2016). In addition, the more data is available, the stronger the association and prediction accuracy of behaviour (Azucar et al., 2018).

Thirdly, the ratings by our human raters might have been influenced by the raters' own anxiety or personality disposition. Even though we used a machine learning approach, it is possible that while learning how the human raters have rated each tweet, the machine learning algorithm also picked up the raters' personality. This point is vital and needs to be explored in future research. However, we are confident that this is potentially only a small influencing factor. That is due to two reasons: a) each tweet was evaluated by several raters, and b) the sets of raters for each tweet are composed randomly. Hence, in merely a few instances would an exact set of raters (5-6 per tweet) evaluate multiple tweets, due to the randomized display of tweets, and the randomized rater allocation to each tweet.

Fourthly, anxious individuals are more likely to disclose intimate information about themselves to others to seem more approachable (Collins & Miller, 1994). Hence, they also use social media to learn about others (Seidman, 2013). They are more likely to use Twitter to do so, since most information on Twitter is public. Hence, it is possible that our results might differ if we had chosen Facebook as our data source. However, based on previous research (Azucar et al., 2018; Settanni et al., 2018) it is unlikely that social media type (public vs. private) would influence the prediction accuracy of the psychological characteristic anxiety.

Finally, one could argue that culture plays an important role as well. As has been noted in previous studies (Masuda & Nisbett, 2001; Tskhay & Rule, 2014), there is likely a difference in the ratings between participants who come from East Asian cultures rather than Western cultures. We are confident to have avoided this pitfall, since we included only participants as raters who are from the United States. Tweets were subsequently also screened for geographical location, based on location and time zones, to ensure that American participants rate tweets only from American (or in the US residing) tweeters.

**Conclusion**

In this study, we have employed a machine learning model to scale the ability to score more than three million tweets on a perceived anxiety scale. The method is non-invasive and longitudinal providing fine-grained anxiety scores for 1,418 tweeters timelines, spanning 18 months on average per user. Data analysis reveals that perceived state anxiety fluctuates per each user over time, but some users have a higher long term mean trait anxiety than others do. Data analysis also showed a statistically significant, reverse relationship between trait anxiety and users' social engagement and popularity, respectively. The presented method can have practical implications in organizations as it allows the automatic monitoring of employees' wellbeing and could be used to prevent the onset of burnout, benefitting workers, organizations, and society alike.

**Table 1 – Correlations among variables**

|   | | M | SD | 1 | 2 | 3 | 4 | 5 |
|---|---|---|---|---|---|---|---|---|
| 1 | Anxiety | 2.34 | .34 | (.90) | | | | |
| 2 | Follower Count | 4483.88 | 64562.07 | -.01*** | | | | |
| 3 | Following Count | 1221.43 | 4235.46 | -.02*** | .08*** | | | |
| 4 | Tweet Count | 35697.49 | 75479.47 | .03*** | .04*** | .29*** | | |
| 5 | Tweet Reply | .18 | .38 | .03*** | -.02*** | -.03*** | -.04*** | |
| 6 | Retweet | .53 | .49 | .04*** | -.01*** | -.02*** | .03*** | -.49*** |

*** $p < .001$, n = 3,292,193 observations

**Table 2 – Impact of Perceived Anxiety on Popularity (Follower Count) and Social Engagement (Following Count) by User**

| | Popularity (Follower Count) | | | | Social Engagement (Following Count) | | | |
|---|---|---|---|---|---|---|---|---|
| | *6 months* | | *12 months* | | *6 months* | | *12 months* | |
| | M1 | M2 | M1 | M2 | M1 | M2 | M1 | M2 |
| Anxiety | -.60*** | -.79 | -.58* | -.77** | -.44† | -.69 | -.39 | -.64* |
| | (-2.72) | (-3.22) | (-2.35) | (-2.85) | (-1.64) | (-2.54) | (-1.43) | (-2.34) |
| Tweet Count | | .00*** | | .00*** | | .00*** | | .00*** |
| | | (4.02) | | (4.04) | | (4.42) | | (4.43) |
| const | 9.56*** | 9.86*** | 9.50*** | 9.81 | 7.93*** | 8.33*** | 7.80*** | 8.22*** |
| | (16.75) | 17.85 | (15.69) | (16.56) | (12.04) | (12.31) | (11.73) | (12.01) |
| Pseudo R² | | .034*** | | .034*** | | .16*** | | .16*** |

n = 1322 users; † $p < .10$, *** $p < .001$, ** $p < .01$, * $p < .05$; z-statistics in parentheses.

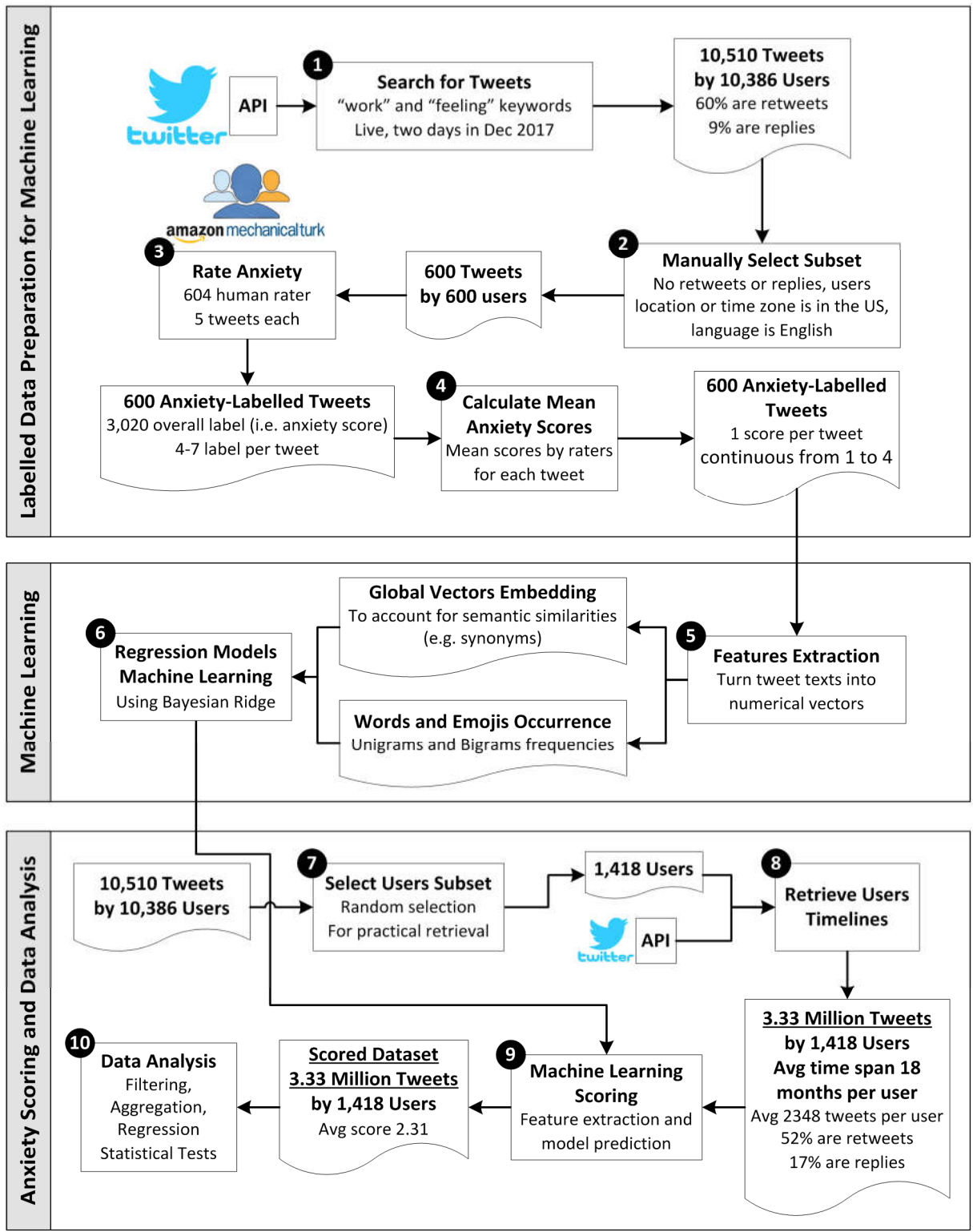

**Figure 1 – Research (Machine Learning) Methodology**

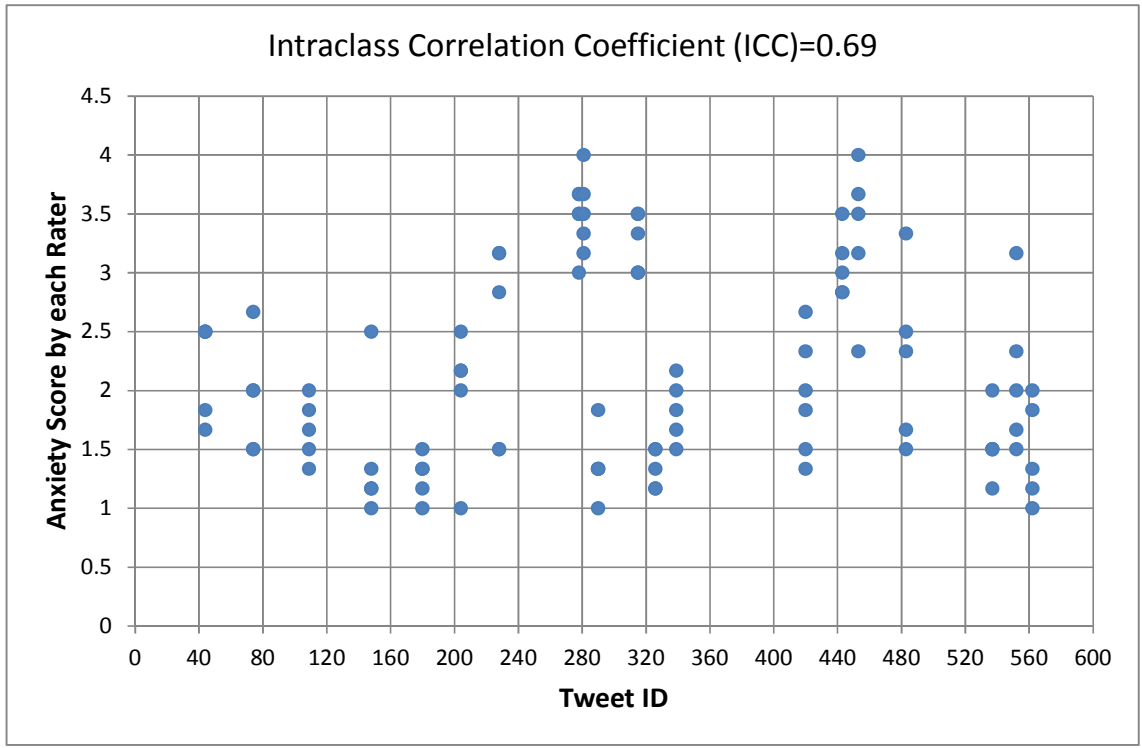

**Figure 2 – Interclass Correlation for Randomly Selected 20 Tweets.**

Each dot in the dot plot represents a rating form a particular rater for the corresponding tweet. Clustered dots for a tweet represent high inter-rater agreement. A random subset of tweets are selected for visualization purposes.

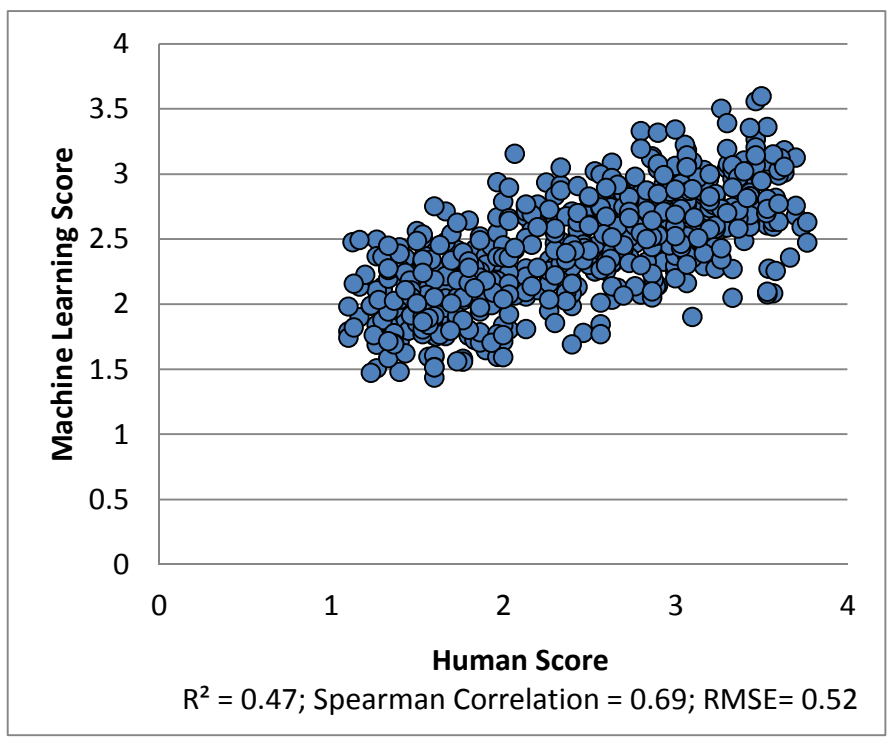

$R^2$ = 0.47; Spearman Correlation = 0.69; RMSE= 0.52

**Figure 3 – Correlation between Machine Learning Anxiety Scores and Human Scores**

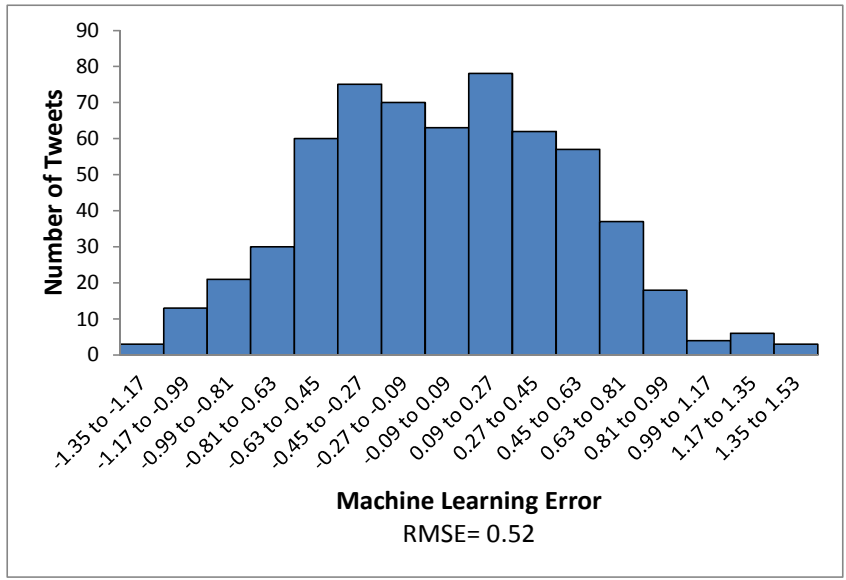

**Figure 4 – Histogram of Machine Learning Error**

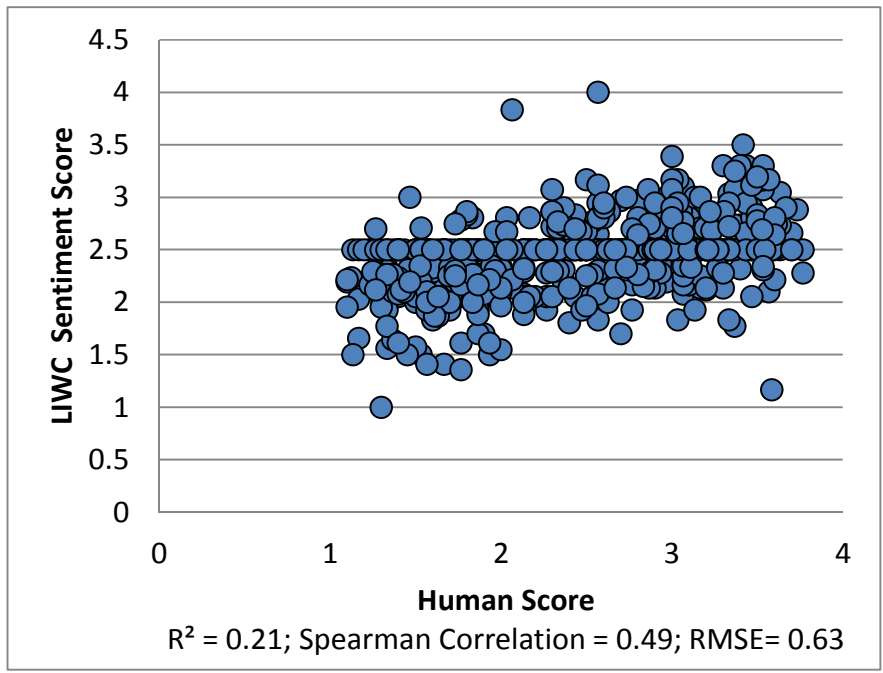

**Figure 5 – Correlation between LIWC Normalized Sentiment Scores and Human Scores**

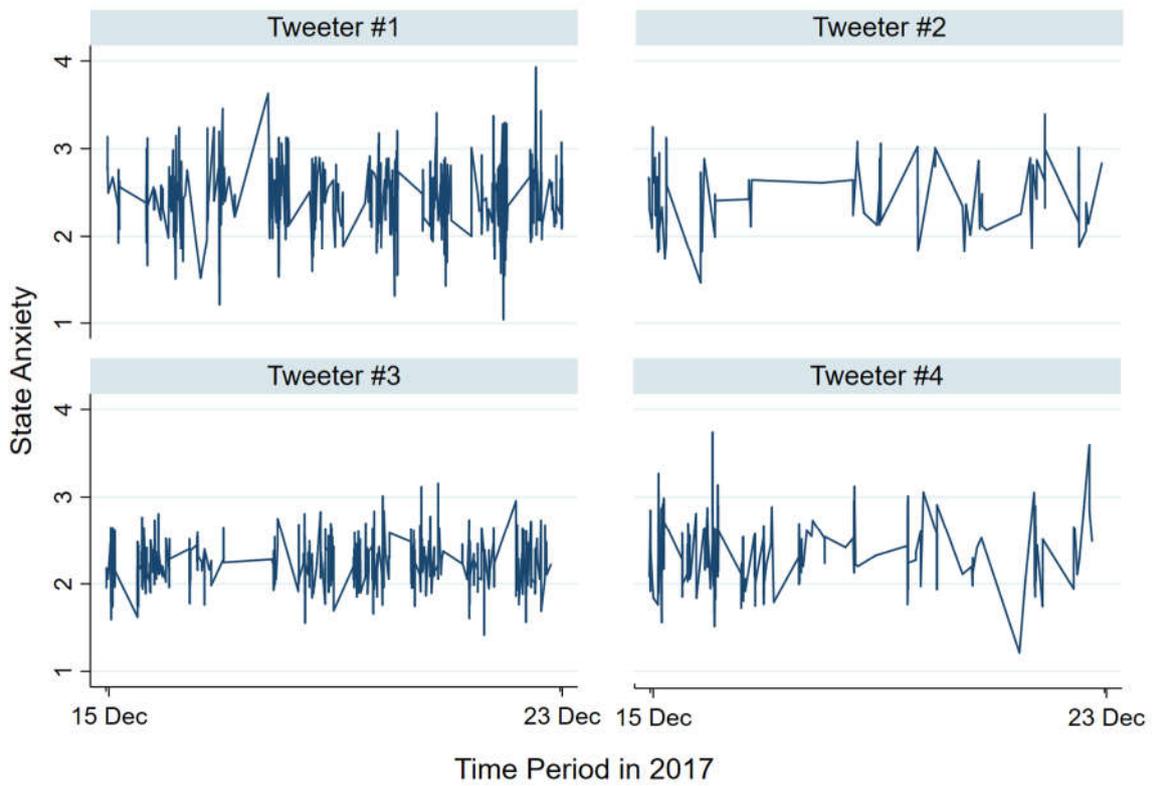

**Figure 6 – Visualization of Random Tweeters' Anxiety Scores over Time**